\documentclass{emulateapj}
\newcommand{\kms}   {km~s$^{-1}$}
\newcommand{\persec}{s$^{-1}$}
\newcommand{\mpy}   {mas~yr$^{-1}$} 
\newcommand{\pmra}  {\mu_{\alpha}}

\newcommand{\pduck} {B1757$-$24}
\newcommand{\pctb}  {B1951$+$32}

\shorttitle{Proper Motions of PSRs B1757$-$24 and B1951$+$32}
\shortauthors{Zeiger et al.}
\slugcomment{ApJ, accepted}

\begin{document}
\title{Proper Motions of PSRs B1757$-$24 and B1951$+$32:\\Implications
for Ages and Associations}

\author{B. R. Zeiger\altaffilmark{1}, W. F. Brisken\altaffilmark{2},
  S. Chatterjee\altaffilmark{3}$^,$\altaffilmark{4}, W. M. Goss\altaffilmark{2}}
\altaffiltext{1}{Center for Astrophysics and Space Astronomy,
  University of Colorado, Boulder, CO 80304; zeigerb@colorado.edu} 
\altaffiltext{2}{National Radio Astronomy Observatory, 
    Socorro, NM 87801; wbrisken, mgoss@aoc.nrao.edu}
\altaffiltext{3}{School of Physics, The University of Sydney, NSW
   2006, Australia; schatterjee@usyd.edu.au}
\altaffiltext{4}{Jansky Fellow}

\begin{abstract}
Over the last decade, considerable effort has been made to measure the
proper motions of the pulsars B1757$-$24 and B1951$+$32 in order to
establish or refute associations with nearby supernova remnants and to
understand better the complicated geometries of their surrounding
nebulae.  We present proper motion measurements of both pulsars with
the Very Large Array, increasing the time baselines of the
measurements from 3.9~yr to 6.5~yr and from 12.0~yr to 14.5~yr,
respectively, compared to previous observations.  We confirm the
nondetection of proper motion of PSR~B1757$-$24, and our measurement
of ($\mu_{\alpha}, \mu_{\delta}) = (-11 \pm 9, -1 \pm
15)$~mas~yr$^{-1}$ confirms that the association of PSR~B1757$-24$
with SNR~G5.4$-$1.2 is unlikely for the pulsar characteristic age of
15.5~kyr, although an association cannot be excluded for a
significantly larger age.  For PSR~B1951$+$32, we measure a proper
motion of ($\mu_{\alpha}, \mu_{\delta}) = (-28.8 \pm 0.9, -14.7 \pm
0.9)$~mas~yr$^{-1}$, reducing the uncertainty in the proper motion by
a factor of 2 compared to previous results.  After correcting to the
local standard of rest, the proper motion indicates a kinetic age of
$\sim$51~kyr for the pulsar, assuming it was born near the geometric
center of the supernova remnant.  The radio-bright arc of emission
along the pulsar proper motion vector shows time-variable structure,
but moves with the pulsar at an approximately constant separation
$\sim$2.5\arcsec, lending weight to its interpretation as a shock
structure driven by the pulsar.
\end{abstract}


\section{INTRODUCTION}
Neutron stars (NSs) are born from the core collapse of massive stars, whose
deaths are marked by supernova remnants (SNRs).  Measuring the proper
motions of young NSs allows their trajectories to be
traced back, and can thus provide strong tests of proposed NS-SNR
associations.  In cases where such associations can be confirmed, an
independent age estimate can be derived for both the NS and its associated
SNR.

However, such an exercise is subject to several pitfalls.  On one
hand, a typical radio pulsar might be detectable for $\gtrsim
10^7$~yr, while the associated SNR fades into the interstellar medium (ISM)
in $\lesssim 10^5$~yr, leaving behind young pulsars with no detectable
SNRs.  On the other hand, there are also several SNRs with no
associated pulsars.  Not all NSs are radio pulsars, but when young,
they are all likely to be bright in thermal X-ray emission, as
illustrated, for example, by the detection of point sources in the SNRs Kes 73 \citep{vg97} and Kes 79 \citep{ssss03}.
The absence of such X-ray detections in a set of nearby,
young SNRs \citep{kfg+04,kgks06} aggravates the problem.  While
unusual cooling scenarios may be required for NSs in these young SNRs,
a partial explanation for the ``missing'' NSs might lie in their birth
velocities.  Radio pulsars have characteristic birth velocities $\sim$
400-500~\kms\ \citep{acc02,hllk05}, and they are likely to escape
their natal remnants once the expansion of the SNR is decelerated by
the ISM, making identification of pulsar-SNR
pairs problematic.

Thus, cases where a young radio pulsar can be associated with a SNR
are particularly important.  The pulsars B1757$-$24 and B1951$+$32
present two such situations in which associations might be possible.
Both are young pulsars with high spin-down energy-loss
rates, and in both cases, the ram pressure balance between the NS
relativistic wind and the ISM produces a pulsar wind nebula (PWN) 
with a bow shock structure.  PSR~B1757$-$24 appears to be exiting the
approximately circular SNR~G5.4$-$1.2, whose asymmetric brightness, 
in conjunction with the PWN produced by the pulsar, produces the
structure known as ``the Duck'' (see Fig.~\ref{fig:duck}).
Meanwhile, PSR~B1951+32 drives a complex interaction within the
SNR~CTB~80 (G69.0$+$2.7; see Fig.~\ref{fig:ctb80}), including a
radio-bright structure resembling a bow shock in the direction of its
motion.

We present Very Large Array (VLA) observations that improve the time
baseline for the proper motion measurements of these two pulsars.  In
\S\ref{sec:duck} we detail the observations, analysis, and results
for PSR~B1757$-$24, and likewise for PSR~B1951$+$32 in \S\ref{sec:ctb80}.
We discuss the implications of our results in \S\ref{sec:discuss}.

\section{PSR~B1757$-$24: OBSERVATIONS AND RESULTS}
\label{sec:duck}
\begin{figure*}[thb]
\epsscale{1.0}
\plotone{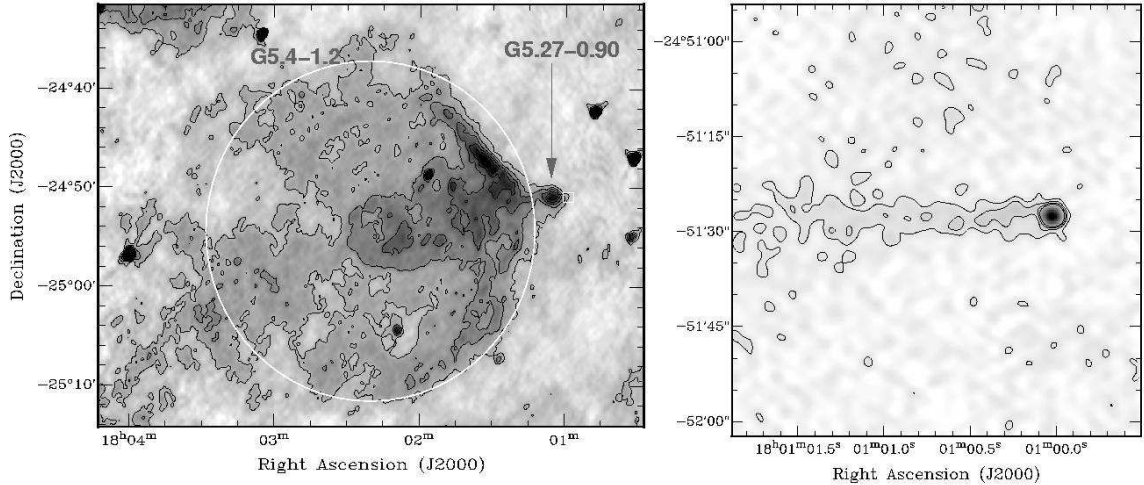}
\caption{The Duck: the SNR G5.4$-$1.2, G5.27$-$0.90, and \pduck\
  system. {\it Left}: Large-scale structure of the remnant, as observed by
  the VLA at 327~MHz with an angular resolution of $40\arcsec \times
  23\arcsec$ \citep[image data courtesy of the authors]{bgg+06}.  The
  extended circular structure is SNR~G5.4$-$1.2, while G5.27$-$0.90 is
  the ``head'' of the Duck protruding from the west edge of the
  remnant.  The small white square shows the approximate location of
  the right-hand panel.
  {\it Right}: Pulsar \pduck\ and its PWN (the western tip of
  G5.27$-$0.90), as observed at 1.4~GHz in 2002.  The large-scale
  structures are resolved out by the interferometer response, leaving
  only the ``beak'' visible at an angular resolution of $2\farcs2
  \times 1\farcs2$.
\label{fig:duck} 
}
\end{figure*}

PSR~\pduck\ is an energetic young pulsar, with a period $P = 125$~ms,
a spindown energy loss rate $\dot{E} = 10^{36.4}$~erg~\persec, and a
characteristic age $\tau_c \equiv P/2\dot{P} = 15.5$~kyr.  As shown in
Figure~\ref{fig:duck}, its location and the morphology of its PWN
suggest that it is escaping the circular SNR~G5.4$-$1.2
\citep{fk91,mkj+91}.  If it was born at the center of the remnant and
its characteristic age is close to its real age, that would imply a
proper motion $\sim$ 70~\mpy\ and a transverse velocity
$\sim$1500-2000~\kms, high compared to the radio pulsar population. 
However, \citet{gf00} placed an upper limit on the motion of the PWN
that was inconsistent with such a high velocity, and suggested that
the true age of the pulsar was much larger than the characteristic
age.  \citet{tbg02} derived comparable limits on the proper
motion\footnote{All uncertainties are 68\% confidence intervals,
except where both 68\% and 95\% intervals are explicitly reported.}
of the pulsar (rather than the PWN) of ($\mu_{\alpha}, \mu_{\delta}) =
(-2.1\pm7.0, -14\pm13)$~mas~yr$^{-1}$ and suggested instead that the
proximity of B1757$-$24 and G5.4$-$1.2 was merely a line-of-sight
coincidence in the crowded region near the Galactic center.  Arguments
for and against an association between the pulsar and SNR were
summarized by \citet{bgc+06}, who improved the upper limit on the
westward proper motion of the PWN and concluded that both
interpretations for the Duck (a line-of-sight coincidence and a
genuine association with a large age) remained viable.

Based on the dispersion measure, PSR~B1757$-$24 is estimated to lie at
a distance of 5.2$\pm$0.5~kpc \citep{cl02}, while the distance to
G5.27$-$1.2 is greater than 4.3~kpc based on \ion{H}{1} absorption
\citep{fkw94,tbg02}.  Here we adopt a distance of $5d_5$~kpc.
\subsection{Observations}
\label{sec:1757obs}
VLA observations of PSR B1757$-$24 spanned 6.4~yr, from 1998 to 2004.
The 1998 June 03 epoch was observed using the BnA configuration;
all of the later epochs used the A configuration.  At the frequency of
observation, 1.45~GHz, the A configuration resolution at a declination
of $-25$\arcdeg\ is $\sim$1.2\arcsec$\times$2.2\arcsec, elongated
north-south.  The first three epochs employed the pulsar gate on the
right circular polarization to allow increased sensitivity in the
pulsar measurements by only accepting data when the pulsar was ``on.''
The 2002 February 21, 2002 May 02, and 2004 December 09 epochs included the Pie
Town VLBA antenna, more than doubling the maximum east-west baseline
length.  The ungated data from the 2002 February 21 epoch was not used, as
the observation was not long enough to provide sufficient {\it u-v}
coverage and sensitivity needed for clear identification of the
pulsar.  In order to minimize variation in {\it u-v} coverage between epochs and in contrast to \citet{tbg02}, we do not use the Pie Town
baselines.
Observations were made in the `2AD' correlator mode, offering
15 spectral channels each of width 1.56~MHz in both circular
polarizations and allowing a usable field of view of $\sim$30\arcmin.
Details of these observations are shown in Table~\ref{tab:obs-duck}.

The data were calibrated using standard procedures within the AIPS\footnote{See http://www.aips.nrao.edu/}
package.  Sources 3C~286 and 3C~48 were used as flux density and
bandpass calibrators and 1751$-$253 was used for phase calibration.
The AIPS task UVFIX was used to recalculate the {\it UVW} coordinates
incorporating aberration corrections not performed by the VLA online
system.  Given the large field of view and the sparse distribution of
sources with compact structure, we selected 14 sub-regions of the
field (containing the pulsar and 13 other apparently compact sources)
for imaging.  These fields were jointly deconvolved to produce initial
images.  Two self-calibration and imaging iterations were performed
using all 14 sub-regions to generate the final images used for the
astrometry.

Positions of reference sources and the pulsar were determined with a
Gaussian fit using the AIPS task JMFIT.  The proper motion of the
pulsar was measured with respect to a reference frame defined by six
point sources chosen based on compactness from the 13 imaged
sub-regions.  The reference source positions and fluxes from the
(ungated) 2004 December 09 epoch are listed in Table~\ref{tab:refsrc-duck},
and our method is described in greater detail in \citet{mcgary}.

The 10\arcsec\ PWN in which the pulsar is embedded is resolved out by
our observations, but it contributes an additional uncertainty to the
position of the pulsar in the ungated data.  To account for the
$\sim$50\% increase in rms noise within this region relative to the
background, the pulsar position uncertainties were increased in
ungated epochs by a factor of 1.5.  In the gated data, the PWN
contribution to the noise is insignificant and the position
uncertainties of the reference sources dominate the proper motion
uncertainty.
\begin{deluxetable*}{ccccccccc}\vspace{-6mm}
  \tablecolumns{9}
  \tablewidth{0pc}
  \tablecaption{Observational Parameters for B1757$-$24\label{tab:obs-duck}} 
  \tablehead{
    \colhead{Epoch} &
    \colhead{Obs.}&
    \colhead{Poln.}&
    \colhead{Gated}&
    \colhead{Flux density}&
    \colhead{rms noise}&
    \colhead{Beam}&
    \colhead{Pos.}&
    \colhead{T$_{\rm int}$} \\
    \colhead{} &
    \colhead{Code} &
    \colhead{}&
    \colhead{}&
    \colhead{(mJy)}&
    \colhead{(mJy beam$^{-1}$)}&
    \colhead{(\arcsec)}&
    \colhead{angle}&
    \colhead{(hr)}
  }
  \startdata 
1998 Jun 03 & AF336  & R & Y & 5.0 & 0.3 & 2.3$\times$1.2 & 16\arcdeg & 1.7 \\
            &        & L & N & 1.9 & 0.1 & 2.2$\times$1.2 & 16\arcdeg & 1.7 \\
2001 Jan 01 & AB969  & R & Y & 3.4 & 0.2 & 2.3$\times$1.2 & $-$1\arcdeg & 3.2 \\
            &        & L & N & 1.1 & 0.1 & 2.2$\times$1.2 &  0\arcdeg & 3.2 \\
2002 Feb 21 & AB1029 & R & Y & 3.4 & 0.2 & 2.0$\times$1.2 &  5\arcdeg & 1.5 \\
2002 May 02 & AB1029 & I & N & 4.5 & 0.1 & 2.2$\times$1.2 &  4\arcdeg & 2.3 \\
2004 Dec 09 & AB1139 & I & N & 1.3 & 0.1 & 2.3$\times$1.2 &  0\arcdeg & 3.5
\enddata
\tablecomments{All observations are at a frequency of 1.45~GHz. Pulsar
gating improves the apparent pulsar flux density by a factor $\lesssim [(T_{\rm on} + T_{\rm off})/T_{\rm on}]^{1/2}$.}
\end{deluxetable*}

\begin{deluxetable}{cccc}
\tablecolumns{5}
\tablewidth{0pc}
\tablecaption{Reference Sources Used in the Proper Motion Fit of \pduck\label{tab:refsrc-duck}} 
\tablehead{
\colhead{R.A.}  &
\colhead{Decl.}  &
\colhead{Flux density} &
\colhead{$\theta_{\mathrm{sep}}$}\\
\colhead{(J2000)}&
\colhead{(J2000)}&
\colhead{(mJy beam$^{-1}$)}&
\colhead{(\arcmin)}
}
\startdata
18 01 27.40 & $-$25 07 38.1 & 8.7(2) & 17.3 \\
18 01 58.57 & $-$24 55 48.5 & 6.6(2) & 14.0 \\
18 00 47.31 & $-$24 42 45.0 & 2.6(1) & 9.2 \\
18 00 15.71 & $-$25 00 22.0 & 3.6(1) & 13.4 \\
18 01 28.11 & $-$24 34 23.7 & 1.2(2) & 18.2 \\
18 00 41.14 & $-$24 42 04.4 & 4.9(1) & 10.3 \\
\enddata
\enddata
\tablecomments{Units of right ascension are hours, minutes, and seconds, and units of declination are degrees, arcminutes, and arcseconds.}
\end{deluxetable}

\subsection{Results}
The structure of the PWN around the pulsar is uncertain and could
cause a systematic offset in the measurement of the pulsar position,
affecting registration between the gated and ungated data sets.  Thus,
we measure the proper motion of the pulsar separately for the gated
and ungated data and combine the two independent results for a final
proper motion value.  Using ungated data, we measure a proper motion
of $(\mu_{\alpha}, \mu_{\delta}) = (-16.9\pm12.6,
-35.8\pm22.4)$~mas~yr$^{-1}$; with gated data, we measure
$(\mu_{\alpha}, \mu_{\delta}) = (-5.9\pm11.8,
6.0\pm20.3)$~mas~yr$^{-1}$. Combining the results gives a measurement
of $(\mu_{\alpha}, \mu_{\delta}) = (-11\pm9,
-1\pm15)$~mas~yr$^{-1}$, consistent with a nondetection.  Note that
correction to the local standard of rest (LSR) for the pulsar
involves adjustments at the level of 1~mas~yr$^{-1}$. Since these are
much lower than the uncertainties, we ignore these corrections for
B1757$-$24. 

We note that our results and those of \citet{tbg02} are not
independent, since two epochs of data are common
between the two analyses, although the Pie Town link data were not
used in the present work.  The measurement of \citet{bgc+06} is
completely independent, since they observe the nebula and not the
pulsar. However, only one dimension of motion was probed by these
authors.  We derive a 68\% confidence upper limit on the {\em
westward} motion of the pulsar, $\pmra > -14.9$~\mpy\ ($\pmra >
-24.8$~\mpy\ at 95\% confidence), which corresponds to a westward
transverse velocity $v_\perp < 360 d_5$~\kms\ ($v_\perp < 600
d_5$~\kms\ at 95\% confidence).  The implications are discussed
further below, but we note that such velocity limits are 
consistent with the velocity distribution of ordinary young radio
pulsars \citep{acc02,hllk05}.

\section{PSR~B1951$+$32: OBSERVATIONS AND RESULTS}
\label{sec:ctb80}

PSR~B1951$+$32, a 39.5~ms pulsar with $\tau_c=107$~kyr and a spin-down
energy-loss rate $\dot{E}=10^{36.6}$~erg~s$^{-1}$, lies near the
southwestern edge of the approximately circular infrared shell of
CTB~80 \citep{fesen}.  The pulsar is surrounded by a
$\sim$30\arcsec\ diameter asymmetric PWN at the western edge of an
8\arcmin$\times$4\arcmin\ east-west plateau of emission
\citep{castel03}.  CTB~80, observed at 1.4~GHz to have three arms
that cover 1.8~deg$^2$ and converge near B1951$+$32 \citep{castel03},
has an expanding \ion{H}{1} shell that yields a dynamical age estimate
of 77~kyr for the pulsar-SNR system \citep{Koo}.  From its dispersion
measure, the distance to B1951$+$32 is estimated as $3.1 \pm 0.2$~kpc
\citep{cl02}, while the distance to CTB~80 is estimated as
2~kpc from \ion{H}{1} absorption \citep{ss00}. Here we adopt a
distance of $2d_2$~kpc.  \citet{kulk88} estimated a speed of
300~km~s$^{-1}$ for the pulsar from scintillation measurements, while
\citet{mig02} directly measure a proper motion $\mu = 25 \pm 4$~\mpy\
at a position angle $252\arcdeg \pm 7\arcdeg$ after correcting for the
effects of differential Galactic rotation. Their implied transverse 
velocity of the pulsar is $V_\perp = (240 \pm 40)\;d_2$~\kms.

\begin{figure*}[th]
\epsscale{0.9}
\plotone{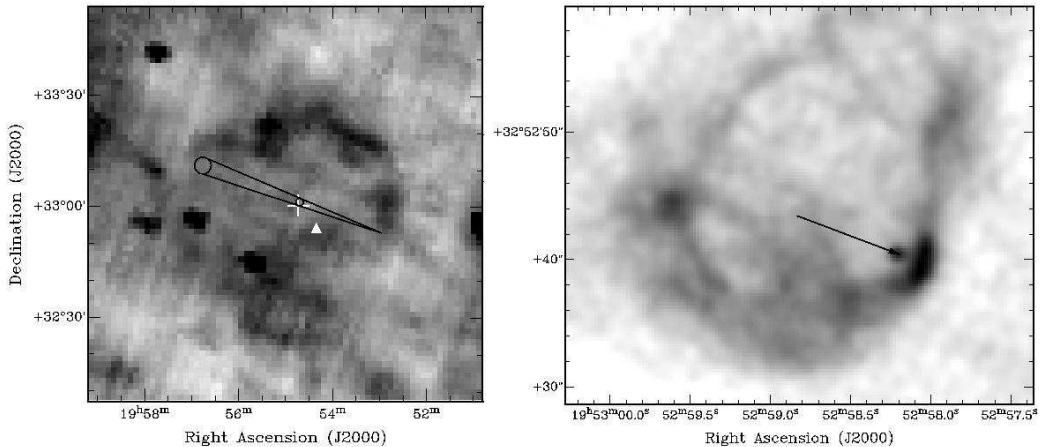}\vspace{-2mm}
\caption{CTB~80 (G69.0$+$2.7) and the PWN produced by PSR~\pctb.
{\it Left}: Infrared ratio image (60$\mu$m/100$\mu$m) constructed from
the {\it IRAS} archive shows the nearly complete shell of
CTB~80. Overplotted lines indicate the projected path of the
pulsar, projected back in time with the 1 $\sigma$ uncertainty on the
measured proper motion, corrected to the LSR.
Circles indicate the estimated birth locations for age estimates
of $\tau_p \sim 51$~kyr (the age of closest approach to the SNR
geometric center, as shown with the white cross) and the
characteristic (spin-down) age $\tau_c = 107$~kyr.  The projected path
does not pass close to the expansion center obtained by \citet{Koo},
indicated by the white triangle. {\it Right}: VLA 1.4~GHz image of
B$1951+32$ and its PWN, with an angular resolution of
1\farcs0$\times$0\farcs9 and rms noise 0.03~mJy~beam$^{-1}$.  The arrow
represents 300~yr of travel along the proper motion vector (see
\S\ref{sec:1951results}). The bow shock is visible as the bright arc
to the south-west of the pulsar.  Note the difference in scale between
the two panels.
\label{fig:ctb80} 
}
\end{figure*}  

In the direction of the proper motion vector measured by
\citet{mig02}, \citet{moon04} observe a cometary X-ray synchrotron
nebula and an H$\alpha$ bow shock.  The X-ray emission peaks at the
pulsar location and is confined within the H$\alpha$ structure
\citep{hester00} that is
clearly defined at an angular separation of $\sim$7\arcsec\ from the
pulsar.  The X-rays are produced by synchrotron emission from the
pulsar wind, confined within a bow shock produced by the ram pressure
of the wind interacting with the SNR wall.

In radio images, B1951+32 appears to be situated just inside a
limb-brightened bubble $\sim$30\arcsec\ in diameter (see
Fig.~\ref{fig:ctb80}, {\it right}).  A 5\arcsec\ portion of the
bubble nearest to the pulsar is substantially brighter than any other
portion of the shell. In MERLIN observations at 1.6~GHz with 0.15\arcsec\
resolution, \citet{golden05} find that the radio-bright arc shows
compact structure, resembling a radio bow shock.  The structure is
$\sim$2.5\arcsec\ from the pulsar, which puts it at the head of the
cometary X-ray nebula, but enclosed within the H$\alpha$ bow shock
which is $\sim$7\arcsec\ away from the pulsar \citep{moon04}.

\subsection{Observations}
\label{sec:1951obs}
Measurements of the proper motion of PSR~B1951$+$32 were made using five
epochs of archival VLA data from 1989.04 to 2003.50. Observation
details are shown in Table~\ref{tab:obs-ctb80}. All observations were
made with the VLA A~array except the 2000 November 25 epoch, which included
the Pie Town antenna. To maintain consistent {\it u-v} coverage in
each epoch, the baselines to Pie Town were removed before data were
reduced.  Observations were made with 15 channels of 1.56~MHz each.
Data were calibrated with VLA calibrator sources 3C~286 
(for flux density and bandpass calibration) and
J1925$+$211 (for phase calibration), following the method outlined in
\S\ref{sec:1757obs}.  Nineteen sources with compact structure were
found in the field of view and were included in the imaging and
self-calibration process.  Eight pointlike sources (see 
Table~\ref{tab:refsrc-ctb80}) were used in the determination of the
proper motion in the manner described in \S\ref{sec:1757obs}.

The pulsar B1951+32 lies in a complicated region of the CTB~80
SNR and is embedded in its PWN, making the position fit
sensitive to observation parameters such as observing frequency and
{\it u-v} coverage at each epoch.  To account for this to first order,
a linear brightness gradient in a 1.75\arcsec\ square around the pulsar
was fit simultaneously with the pulsar position.  Trial fits performed
without this procedure resulted in substantially greater scatter in
the position measurements.  Additional trials demonstrated that the
fit was not very sensitive to the exact size and position of the
region fitted.

\begin{deluxetable*}{cccccccccc}\vspace{-6mm}
  \tablecolumns{10}
  \tablewidth{0pc}
  \tablecaption{Observational Parameters for B1951$+$32\label{tab:obs-ctb80}}
  \tablehead{
    \colhead{Epoch} &
    \colhead{Obs.}&
    \colhead{IF}&
    \colhead{Pol.}&
    \colhead{Freq.}&
    \colhead{Flux density}&
    \colhead{rms noise}&
    \colhead{Beam}&
    \colhead{Position}&
    \colhead{T$_{int}$} \\
    \colhead{} &
    \colhead{Code} &
    \colhead{}&
    \colhead{}&
    \colhead{(MHz)} &
    \colhead{(mJy)}&
    \colhead{(mJy bm$^{-1}$)}&
    \colhead{(\arcsec)} &
    \colhead{Angle}&
    \colhead{(hr)}
  }
  \startdata
  1989 Jan 13 & AS357 & 1 & RR & 1385 & 1.01 & 0.07 & 1.1$\times$1.1 &  
50\arcdeg & 5.9 \\
              &       & 2 & LL & 1652 & 0.62 & 0.09 & 1.0$\times$0.9 &  
81\arcdeg & 5.9 \\
  1991 Jul 18 & AF214 & 1 & RR & 1385 & 1.95 & 0.08 & 1.3$\times$1.2 & $-
$47\arcdeg & 5.8\\
              &       & 2 & LL & 1652 & 1.32 & 0.08 & 1.0$\times$0.9 & $-
$74\arcdeg & 5.8\\
  1993 Jan 08 & AF235 & 1 & RR & 1385 & 1.75 & 0.06 & 1.2$\times$1.2 & $-
$64\arcdeg & 5.9\\
              &       & 2 & LL & 1652 & 1.79 & 0.05 & 1.1$\times$1.0 & $-
$58\arcdeg & 5.9\\
  2000 Nov 25 & AG602 & 1 & RR,LL & 1385 & 0.63 & 0.05 & 1.3$\times$1.2 & $-
$69\arcdeg & 5.2\\
              &       & 2 & RR,LL & 1516 & 0.91 & 0.07 & 1.1$\times$1.0 & $-
$75\arcdeg & 5.2\\
  2003 Jun 03 & AG650 & 1 & RR,LL & 1665 & 1.05 & 0.03 & 1.0$\times$0.9 & $-
$70\arcdeg & 6.2\\ 
  \enddata
\end{deluxetable*}

\begin{deluxetable}{cccc}\vspace{-5mm}
\tablecolumns{4}
\tablewidth{0pc}
\tablecaption{Reference Sources Used in the Proper Motion Fit of B1951$+$32\label{tab:refsrc-ctb80}}
\tablehead{
\colhead{R.A.}  &
\colhead{Decl.}  &
\colhead{Flux density} &
\colhead{$\theta_{\mathrm{sep}}$}\\
\colhead{(J2000)}&
\colhead{(J2000)}&
\colhead{(mJy beam$^{-1}$)}&
\colhead{(\arcmin)}
}
\startdata
19 53 23.70 & $+$32 53 35.3 &   4.7(3) & 5.4 \\
19 53 16.37 & $+$32 48 46.5 &  21.7(3) & 5.5 \\
19 53 17.92 & $+$32 49 11.4 &   6.8(3) & 5.4 \\
19 53 13.42 & $+$33 01 21.8 &  97.8(3) & 9.3 \\
19 52 45.25 & $+$33 03 30.0 &   6.2(3) & 11.2 \\
19 52 15.80 & $+$32 49 35.7 & 839.3(8) & 9.4 \\
19 52 30.01 & $+$32 55 26.2 &   5.4(3) & 6.5 \\
19 52 13.32 & $+$32 59 22.1 &  61.7(3) & 11.6
\enddata
\tablecomments{Units of right ascension are hours, minutes, and seconds, and units of declination are degrees, arcminutes, and arcseconds.}
\end{deluxetable}

\subsection{Results}
\label{sec:1951results}

\begin{figure}[h]\vspace{+3mm}
\epsscale{0.9}
\plotone{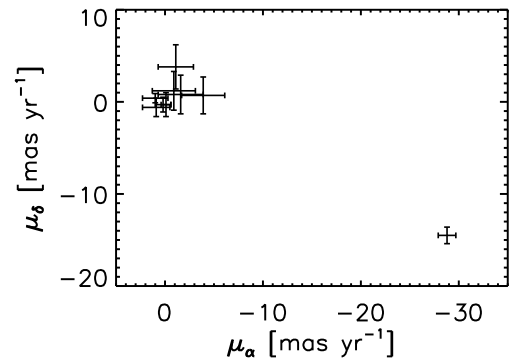}
  \caption{Proper motion of PSR~B1951$+$32 and the reference
   sources used for the fit. Reference source motions with respect to
   each other are consistent with noise with and scattered around
   zero.  The pulsar lies in the bottom right corner with a proper
   motion of $(\mu_{\alpha}, \mu_{\delta})=(-28.8 \pm 0.9,-14.7 \pm
   0.9)$~mas~yr$^{-1}$.  In this figure the derived motion of the
   pulsar is relative to the frame defined by all eight reference
   sources.
\label{fig:mot}
}
\end{figure}

The proper motion of B1951$+$32 was measured as $(\mu_{\alpha},
\mu_{\delta})=(-28.8\pm0.9,-14.7\pm0.9)$~mas~yr$^{-1}$.
Figure~\ref{fig:mot} demonstrates the significance of the proper
motion relative to that of the eight presumably stationary sources
used to define the frame in which the proper motion was measured.  For
each of the reference sources, we derive a proper motion in the frame
defined by the seven other sources. As shown in Figure~\ref{fig:mot},
these proper motion values are consistent with zero at
$\sim$1 $\sigma$, indicating that a self-consistent reference frame has
been established.

A significant correction is required to obtain the motion of the
pulsar in its LSR so that it can be related
to the remnant. We use a flat Galactic rotation curve with velocity
220~km~s$^{-1}$ and assume a solar radius of 8.5~kpc.  Further
correction for the solar peculiar motion is applied \citep{db98}.  The
magnitude of the correction depends on the distance; an additional
0.6~mas~yr$^{-1}$ has been added in quadrature to the proper motion
uncertainty to accommodate the magnitude range of the correction over
the 1--3~kpc distance range considered.  The LSR-corrected proper
motion is found to be $(\mu_{\alpha}^\prime, \mu_{\delta}^\prime)=
(-26.9 \pm 1.1,-10.5 \pm 1.1)$~mas~yr$^{-1}$, for a total proper
motion $\mu^\prime = 29\pm1$~mas~yr$^{-1}$ at a position angle
$249\pm2$\arcdeg\ east of north.

\subsection{Shock Separation}
\label{sec:shock}

To maximize sensitivity to the extended structure of the shock, all
{\it u-v} data within each epoch were added with the AIPS task DBCON.
Before reimaging, the resulting data were convolved with a circular
beam with a 1.2\arcsec\ diameter, the longest semi-major axis of any beam
in any of the concatenated data sets, to ensure equal sensitivity to
large-scale structure across all epochs.

\begin{figure}[htf]
\epsscale{0.9}
\plotone{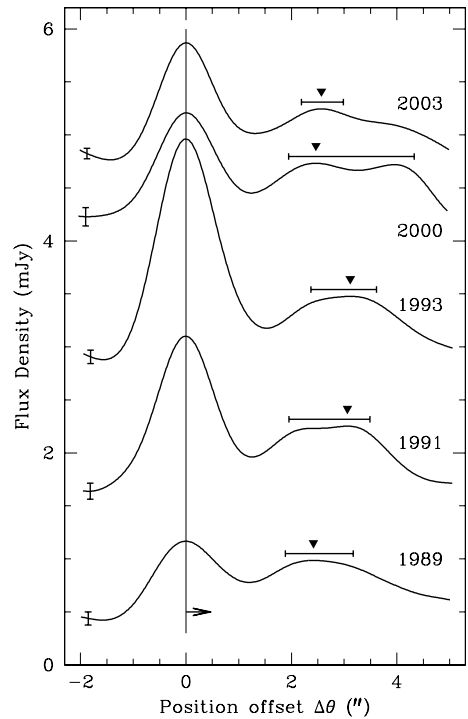}
\caption{Slices through PSR~\pctb\ and its bow shock.  The flux
 density is plotted as a function of the offset from the derived
 pulsar position (as indicated by the solid line), along the measured
 proper motion vector.  Successive epochs are vertically displaced by
 1~mJy (except for 2003, which is displaced by 1.25~mJy) for clarity,
 but note that the time interval between successive observations is
 not uniform.  The map rms noise $\sigma$ at each epoch is indicated
 by the error bar at the far left.  The peak of the shock is marked,
 and the width of the shock within 1 $\sigma$ of the shock peak is also
 shown by the horizontal error bar.  The measured separation is
 consistent with being constant over $\sim$14.4~yr, while the pulsar
 moves $\sim$0.47\arcsec in that time period, as shown by the
 horizontal arrow.
\label{fig:slice} 
}
\end{figure}

Full hydrodynamical modeling of the shock is beyond the scope of the
present work. Instead, we simply extract the flux density in slices
along the observed proper motion vector in the LSR (position angle $=
249\arcdeg$ east of north) in order to estimate the angular separation
between the pulsar and the shock structure at each epoch.  As shown in
Figure~\ref{fig:slice}, the pulsar position is well defined in these
slices, but the transverse structure of the shock changes with epoch
and the separation of the pulsar from the shock peak
($\Delta\theta_{\rm Peak}$) varies between 2.4\arcsec\ and 3.1\arcsec.  At
each epoch, given the map rms noise $\sigma$, we define the shock
thickness as the range where the emission is within 1 $\sigma$ of the
shock peak.  The near and far edges of the shock, defined in this
manner, are listed in Table~\ref{tab:shock} and indicated in
Figure~\ref{fig:slice}.

\begin{deluxetable}{lccc}
  \tablecolumns{4}
  \tablewidth{0pc}
  \tablecaption{Angular Separation between PSR~B1951+32 and the Radio Bow Shock\label{tab:shock}}
  \tablehead{
    \colhead{Epoch} &
    \colhead{$\Delta\theta_{\rm Peak}$}&
    \colhead{$\Delta\theta_{\rm Near}$}&
    \colhead{$\Delta\theta_{\rm Far}$}\\
\colhead{} & \colhead{(\arcsec)} & \colhead{(\arcsec)} & \colhead{(\arcsec)}
}
\startdata
1989 & 2.4 & 1.9 & 3.2 \\
1991 & 3.1 & 2.0 & 3.5 \\
1993 & 3.1 & 2.4 & 3.6 \\
2000 & 2.5 & 1.9 & 4.3 \\
2003 & 2.6 & 2.2 & 3.0 \\
\enddata
\tablecomments{We list the angular distance from the pulsar peak to
    the peak of the shock and the range within 1 $\sigma$ of the shock
    peak at each epoch. The position fit uncertainties are
    insignificant compared to the thickness of the shock itself.}
\end{deluxetable}

Over the 14.5~yr span of observations, the pulsar moves by $\sim$
0.47\arcsec, but the distance between the pulsar and the near edge of
the shock is $\sim$2.1\arcsec$\pm$0.2\arcsec\ and the separation to the
far edge is $\sim$3.6\arcsec$\pm$0.6\arcsec.  Since the separation
between the pulsar and the radio-bright feature does not show a
secular decreasing trend, we concur with previous inferences
\citep{hk88,cc02} that the feature must be a shock driven by the
pulsar wind as the pulsar travels through the ambient medium.

\section{DISCUSSION}
\label{sec:discuss}

\subsection{The Association between SNR~G5.4$-$1.2  and PSR~B1757$-$24}
\label{sec:discussduck}

The identity of the remnant associated by birth with the pulsar
B1757$-$24 is still a mystery.  Based on morphology and location,
SNR~G5.4$-$1.2 is the only known remnant likely to be related to the
pulsar, but that association has serious problems (see Blazek et al. [2006]
for a recent discussion). To remain tenable, such an association
requires a pulsar age much larger than the characteristic age $\tau_c
= 15.5$~kyr, an explosion site much closer to the current location of
the pulsar, or some combination of the two.  For example, G5.4$-$1.2
could be expanding into a region with a density gradient
\citep[e.g.,][]{gvar04}.  The brightening of the remnant on the
western limb, the side nearer to the Galactic plane, may be evidence
for such an asymmetric expansion.  However, even in an optimistic
scenario, the birth site could be no closer than the westernmost limb
of G5.4$-$1.2.  In such a case, for a true age $\le \tau_c$, a
westward proper motion $\gtrsim 18$~\mpy\ is required for an
association, but all three measurements reported in
Table~\ref{tab:summary}, where we summarize our results, imply
westward motions no greater than 15~mas~yr$^{-1}$.  It thus appears
that either the true age is larger than $\tau_c$, probably
significantly so, or the association of the pulsar with G5.4$-$1.2
must be excluded.

We note that the characteristic ages of some young pulsars with
well-determined kinetic ages are overestimates, for example, J1811$-$1925 
\citep{krv+01}, J0538$+$2817 \citep{nrb+07}, and B1951$+$32 \citep[our
\S\ref{sec:discussctb}]{mig02}.  Thus, the true age of B1757$-$24 might
be no greater than 15.5~kyr, rendering the association unlikely.  But
Vela, for example, may have a true age greater than $\tau_c$
\citep{lpg+96}, and \citet{bgc+06} argue that a growing surface
magnetic field, which decouples the true age from the characteristic
age, could account for such an age discrepancy in the Duck while
preserving the association.  Given a long enough time baseline, the
pulsar proper motion (or the motion of the nebula) will surely be
detectable, allowing a firm conclusion about this vexing association.

\begin{deluxetable*}{lccccc}\vspace{0mm}
   \tablecolumns{6} \tablewidth{0pc} 
\tablecaption{Summary of Proper Motion Measurements\label{tab:summary}}
   \tablehead{
	&
	\multicolumn{3}{c}{B1757$-$24}&
	\multicolumn{2}{c}{B1951$+$32}\\
	&
	\colhead{This work} & 
	\colhead{Ref.\ 1} &
	\colhead{Ref.\ 2}&
	\colhead{This work}&
	\colhead{Ref.\ 3}
   }
\startdata
Target                       & B1757$-$24     & B1757$-$24   & G5.27$-$0.9  & 
B1951$+32$   & B1951$+$32 \\
$\mu_\alpha$ (mas yr$^{-1}$) & $-11\pm9$ & $-2.1\pm7.0$&  $>-$7.9      & $-
28.8\pm0.9$& $-29\pm2$  \\
$\mu_\delta$ (mas yr$^{-1}$) & $-1\pm15$ & $-14\pm13$   &  \ldots      & $-
14.7\pm0.9$& $-8.7\pm1.3$\\
Time Baseline (yr)           & 6.5            & 3.9          & 12.0         & 
14.5         & 11.9       \\
Frequency (GHz)              & 1.4            & 1.4          & 8.5          & 
1.4          & 1.4        \\
\\
RA of Pulsar & \multicolumn{3}{l}{\ \ 18 01 00.016$\pm$0.008} & 
\multicolumn{2}{l}{19 52 58.206$\pm$0.001}\\
DEC of Pulsar& \multicolumn{3}{l}{$-$24 51 27.5$\pm$0.2} & 
\multicolumn{2}{l}{32 52 40.51$\pm$0.01}\\
Epoch of RA, DEC&  \multicolumn{3}{l}{\ \ 2004 Dec 09} &\multicolumn{2}{l}{2003 Jun 03}
\enddata
\tablecomments{All provided uncertainties and limits are are 68\% confidence 
   intervals. \citet{bgc+06} do not measure $\mu_\delta$, and their
   68\% $\mu_\alpha$ has been inferred from their 5 $\sigma$ limit; results of 
   \citet{mig02} have been adjusted
   to remove the differential Galactic rotation applied in the published
   result. Proper motions and positions are reported in J2000.0 coordinates
   and do not include LSR corrections. Units of right ascension are hours, minutes, and seconds, and units of declination are degrees, arcminutes, and arcseconds.
}
\tablerefs{(1) \citet{tbg02}; (2) \citet{bgc+06}; (3) \citet{mig02}}
\end{deluxetable*}
\subsection{The Age and Bow Shock of B1951$+$32}
\label{sec:discussctb}

For B1951$+$32, we measure a proper motion of $(\mu_{\alpha},
\mu_{\delta})=(-28.8\pm0.9, -14.7\pm0.9)$~mas~yr$^{-1}$, confirming
and improving on the results of \citet{mig02}.  \citet{mig02} filter
the {\it u-v} plane to remove all spatial scales greater than
4\arcsec\ and to leave only compact sources. We use a different
approach to remove the complexity of the PWN, subtracting the linear
slope of the PWN from the region around the pulsar in the image
domain.  The inclusion of a longer time baseline and a larger field of
reference sources increases both the precision and the accuracy of
this measurement over that of \citet{mig02}.  The new measurement
corresponds to an LSR-corrected transverse velocity of $(274\pm12)\;
d_2$~km~s$^{-1}$.  At a position angle of 249\arcdeg, the proper
motion vector is well matched to the long symmetry axis of the X-ray
PWN \citep{moon04}.

The measurement of the proper motion allows a revised estimate for the
age of the pulsar.  The LSR-corrected proper motion of the pulsar is
shown in Figure~\ref{fig:ctb80}.  In projection, its closest approach
to the expansion center estimated by \citet{Koo} is $\sim$330\arcsec,
a 5.5 $\sigma$ deviation assuming a 1\arcmin\ uncertainty in the
reported expansion center.  A geometric center at coordinates
19h54m50s, 33\arcdeg00\arcmin30\arcsec\ (J2000.0) was estimated by
superposing a circle on the remnant shell seen in the left panel of
Figure~\ref{fig:ctb80}.  The pulsar's closest approach to this point
implies an age $t_p \sim 51$~kyr with a minimum separation of
74\arcsec, consistent with the 56\arcsec\ positional uncertainty based
solely on the proper motion uncertainty.  Our measurement of the
pulsar motion is therefore not consistent with the $\sim$77~kyr age
and the expansion center estimated by \citet{Koo} for the CTB~80 SNR,
but it is consistent with the pulsar age determined by \citet{mig02}. We
note that the true explosion center may be masked by asymmetric expansion
due, for example, to density gradients and structure in the ISM, and
neither estimate for the age may be correct. However, the true age is
almost certainly less than $\tau_c \sim 107$~kyr, since the pulsar
reaches the far edge of the SNR when its measured proper motion is
projected backwards over that time interval, as shown in
Figure~\ref{fig:ctb80}. 

As B1951+32 moves through its ambient medium, its relativistic wind
outflow is confined by ram pressure, and as seen in some other PWNe,
synchrotron emission from the confined pulsar wind produces a cometary
X-ray nebula. On the other side of the contact discontinuity, the
shocked ISM produces H$\alpha$ emission \citep[see,
e.g., the illustration by][]{gvc+04}. In such a picture, the
radio-bright arc would be produced by the confined pulsar wind at the
location of the contact discontinuity.  As we have shown here, the
structure of the bright arc is dynamic, changing from epoch to epoch,
but overall, it moves with the pulsar at an approximately steady
separation $\sim$2.5\arcsec, corresponding to a standoff distance of
$\sim$0.024$\;d_2$~pc.

\subsection{Two Similar Pulsars in Very Different Environments}

Currently, over 1600 radio pulsars have measured values for $P$ and
$\dot{P}$, as listed in the ATNF pulsar
catalog\footnote{See http://www.atnf.csiro.au/research/pulsar/psrcat}
\citep{mhth05}.  In this large ensemble, B1757$-$24 and B1951$+$32 are
among the fifty most energetic pulsars, with $\dot{E} =
10^{36.4}$ and $10^{36.6}$~erg~s$^{-1}$, respectively.
They are also among the hundred youngest known pulsars, with
characteristic ages $\tau_c = 15.5$ and $107$~kyr, respectively.  On one hand, the
characteristic age of B1951$+$32 is likely to be an overestimate, since
its current spin period is not much greater than the typical birth
spin periods of pulsars \citep{fk06}, and its kinematic age $t_p \sim
51$~kyr, assuming birth near the center of CTB~80.  On the other hand,
the upper limits on its westward proper motion suggest that B1757$-$24
may be substantially older than its characteristic age.  Although the true ages of the pulsars differ in relation to their respective characteristic ages,
these two pulsars are quite similar in energetics and age.

Both pulsars also exhibit cometary PWNe.  B1757$-$24 is associated
with a $\sim10$\arcsec\ elongated nebula visible at both radio (as in
Fig.~\ref{fig:duck}) and X-ray wavelengths \citep{kggl01}.  B1951+32
shows a cometary PWN in X-rays \citep{moon04}, but its radio structure
resembles a bow shock and an extended, limb-brightened bubble, and it
is enclosed by a complex nebula visible in H$\alpha$ \citep{hester00},
with extended lobes and a bow shock structure. \citet{hk88} suggest
that the pulsar is interacting with the wall of its evolved SNR,
and we find that the radio shock structure changes with time
while moving with the pulsar.  On the other hand, B1757$-$24 appears
to be outside any parent remnant. If indeed it is associated with
G5.4$-$1.2, it would have interacted with the wall of the remnant in
the past (as B1951+32 is at present), possibly re-energising the shell
and leaving behind the structure we see as the Duck.  Alternatively,
the parent remnant of B1757$-$24 is no longer visible, and we are
witnesses to a cosmic coincidence.

\acknowledgements

B.R.Z. acknowledges support from the Research Experiences for
Undergraduates program of the National Science Foundation and the
Department of Scientific and Academic Affairs of the National Radio
Astronomy Observatory (NRAO) for funding this research. S.C.
acknowledges support from the University of Sydney Postdoctoral
Fellowship program, and he was previously a Jansky Fellow of the
NRAO. The VLA is a facility of the NRAO, funded by the NSF and
operated under cooperative agreement by Associated Universities,
Inc. We thank Dale Frail for helpful conversations regarding this
research, and Crystal Brogan for providing the 327~MHz image of
G5.4$-$1.2.  {\it Facilities:} VLA 

\end{document}